\documentclass[amsmath, amssymb, english, aps, prl, showpacs, twocolumn,  floatfix, superscriptaddress,bibliography]{revtex4-1}

\usepackage[T1]{fontenc}
\usepackage[latin9]{inputenc}

\usepackage[colorlinks=true]{hyperref}  
\hypersetup{
    bookmarks=true,         
    unicode=false,          
    pdftoolbar=true,        
    pdfmenubar=true,        
    pdffitwindow=false,     
    pdfstartview={FitH},    
    pdftitle={Not all doped Mott insulators have a pseudogap: key role of van Hove singularities},    
    pdfauthor={Wu et al.},     
    pdfsubject={},   
    pdfcreator={},   
    pdfproducer={}, 
    pdfkeywords={} {} {}, 
    pdfnewwindow=true,      
    colorlinks=true,       
    linkcolor=blue, 
    citecolor=blue,        
    filecolor=magenta,      
    urlcolor=blue           
}

\usepackage{braket}
\usepackage{bbm}
\newcommand{\equref}[1]{Eq.~(\ref{#1})}

\newcommand{\figref}[1]{Fig.~\ref{#1}}
\newcommand{\refcite}[1]{Ref.~\onlinecite{#1}}

\renewcommand{\vec}[1]{\boldsymbol{#1}}
\newcommand{\pdagger}{{\phantom{\dagger}}}
\renewcommand{\approx}{\simeq}

\usepackage[english]{babel}
\usepackage{graphicx}
\usepackage[usenames,dvipsnames]{color}
\usepackage[normalem]{ulem}
\usepackage{hyperref}
\usepackage{color}

\newcommand{\sect}[1]{\vspace{0.4em}{\it #1.}---}

\begin{document}
\author{Wei Wu}
\affiliation{CPHT, CNRS, \'Ecole Polytechnique, IP Paris, F-91128 Palaiseau, France}
\affiliation{School of Physics, Sun Yat-sen University, Guangzhou, Guangdong Province 510275, China}
\affiliation{Coll\`ege de France, 11 place Marcelin Berthelot, 75005 Paris, France}

\author{Mathias S.~Scheurer}
\affiliation{Department of Physics, Harvard University, Cambridge MA 02138, USA}

\author{Michel Ferrero}
\affiliation{CPHT, CNRS, \'Ecole Polytechnique, IP Paris, F-91128 Palaiseau, France}
\affiliation{Coll\`ege de France, 11 place Marcelin Berthelot, 75005 Paris, France}

\author{Antoine Georges}
\affiliation{CPHT, CNRS, \'Ecole Polytechnique, IP Paris, F-91128 Palaiseau, France}
\affiliation{Coll\`ege de France, 11 place Marcelin Berthelot, 75005 Paris, France}
\affiliation{Center for Computational Quantum Physics, Flatiron Institute,  
162 Fifth avenue, New York, NY 10010, USA} 
\affiliation{DQMP, Universit\'e de Gen\`eve, 24 quai Ernest Ansermet, CH-1211 Gen\`eve, Suisse}

\date{\today}

\begin{abstract}
The Mott insulating phase of the parent compounds is frequently taken as starting point for the underdoped high-$T_c$ cuprate superconductors. In particular, the pseudogap state is often considered as deriving from the Mott insulator.
In this work, we systematically investigate different weakly-doped Mott insulators on the square and triangular lattice to clarify the relationship between the pseudogap and Mottness. We show that doping a two-dimensional Mott insulator does not necessarily lead to a pseudogap phase. 
Despite its inherent strong-coupling nature, we find that the existence or absence of a pseudogap depends sensitively on non-interacting band parameters and identify the crucial role played by the van Hove singularities of the system.
Motivated by a SU(2) gauge theory for the pseudogap state, we propose and verify numerically a simple equation that governs the evolution of characteristic features in the electronic scattering rate.
\end{abstract}

\title{Not all doped Mott insulators have a pseudogap: key role of van Hove singularities}

\maketitle

\sect{Introduction}The complex phenomenology of the cuprate high-$T_c$ superconductors is widely thought of as the consequence of introducing mobile charge carriers in a Mott insulator through doping \cite{Lee2006review}. In the regime of low doping, a pseudogap (PG) state ~\cite{ding1996, timusk1999, normanreview} prevails below an onset temperature $T^*$, which lies above the superconducting dome. 
There are a number of theories proposed to describe the dominant physics in this regime,
such as preformed Cooper pairs~\cite{senthil2009}, various emergent ordered states~\cite{kivelson1998,chakravarty2001,kaminski2002,simon2002}, topological order~\cite{Scheurer2018,scheurer2018orbital,sachdev2019}, and Mottness collapse~\cite{weng1999mean,stanescu2006fermi,phillips2010,zaanen2011mottness,imada2011theory}; 
but, to date, a full consensus has not been reached. On the numerical side, prototypical models such as the two-dimensional Hubbard model~\cite{dagotto1994review,scalapino2007} or the $t$-$J$ model~\cite{jaklic2000,haule2003} have been intensively investigated. Regarding the origin of the PG state, different approaches point to short-range antiferromagntic (AF) correlations and proximity to Mottness, including Quantum Monte Carlo~\cite{bulut1994,preuss1997,wu2017controlling}, exact diagonalization~\cite{jaklic2000}, cluster perturbation theory~\cite{kohno2012}, and cluster extensions of dynamical mean-field theory (DMFT)~\cite{georges1996,maier2005,sakai2009,gull2013,gunnarsson2015,wu2018pseudogap}. This is in line with a recent systematic experimental study~\cite{taillefer2018} indicating that all other instabilities are secondary to AF correlations in the opening of the PG.

\begin{figure}[t]
\begin{centering}
\includegraphics[scale=0.8]{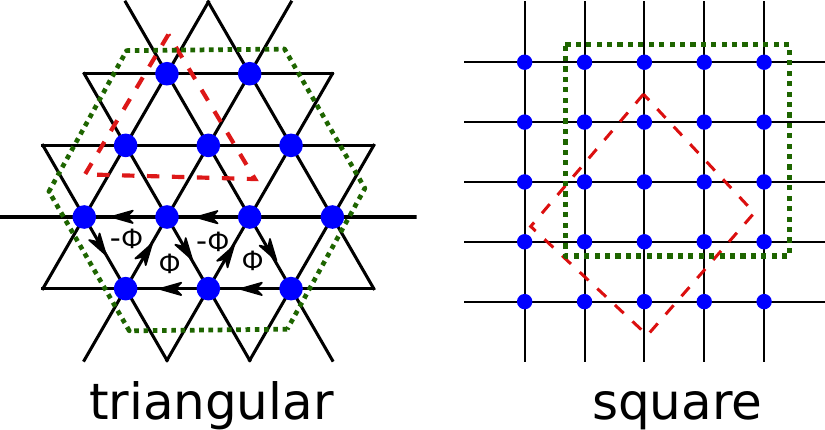}
\par\end{centering}
\caption{Illustration of triangular (left) and square (right) lattice tight-binding models studied in this work. While all hopping parameters are real for the square lattice, we study complex nearest-neighbor hopping on the triangular lattice with $-te^{ + i\Phi/3}$ (along arrow)  or $-te^{ - i\Phi/3}$ (against arrow);
electrons encircling a triangular plaquette once acquire a phase of $\pm\Phi$ as indicated. The geometries of the two different DCA clusters we use for each lattice are indicated by red and green dashed lines.
\label{fig:lattice}}
\end{figure}

Although the importance of  AF correlations and Mottness is stressed by various studies, the exact underlying mechanism of the PG is still an open question. It is natural to expect that \textit{long-range} AF fluctuations can cause a PG due to the coupling of the electrons to collective magnetic modes \cite{PhysRevB.41.6399,PhysRevB.42.7967,refId0,PhysRevLett.93.147004}. In hole-doped cuprates, however, this picture does not apply, as the correlation length is significantly reduced. To shed light on the role played by AF correlations and Mottness, 
we address in this work the following fundamental question:
\textit{do all doped Mott insulators with short-range AF correlations have a PG in two dimensions?} To this end, we systematically study Hubbard models on square and triangular lattices, see \figref{fig:lattice}, which have very different magnetic frustration properties. We discover that, surprisingly, on both lattices, the existence of a PG at finite doping in the strong-coupling regime sensitively depends on the electronic dispersion of the \textit{non-interacting} system. Numerical evidence suggests that this dependence can be well described by a simple relation that involves the location of the van Hove singularity (VHS). In addition, we find a simple equation that captures the evolution of the ``quasipole'' \cite{wu2018pseudogap} in the electronic scattering rate which is related to the depletion of the spectral weight in the PG. This equation agrees with an SU(2) gauge theory~\cite{Scheurer2018,scheurer2018orbital,sachdev2019}, where the PG is the result of a Higgs mechanism, physically corresponding to local AF order with large orientational fluctuations.

\sect{Model and method}We use the dynamical cluster approximation (DCA)~\cite{maier2005}, a cluster extension of single-site DMFT~\cite{georges1996review}, 
 to study the triangular- and square-lattice Hubbard models illustrated in \figref{fig:lattice}. For concreteness, we focus on an on-site Hubbard interaction, $U$, and the corresponding Hamiltonians have the form
\begin{eqnarray}
H= -\sum_{i,j,\sigma}t_{ij}c_{i\sigma}^{\dagger}c^\pdagger_{j\sigma}+U\sum_{i}n_{i\uparrow}n_{i\downarrow}-\mu\sum_{i,\sigma}n_{i\sigma}, \label{HubbardModel}
\end{eqnarray}
where $c^\dagger_{i\sigma}$ are electronic creation operators on site $i$ with spin $\sigma=\uparrow,\downarrow$, $n_{i\sigma} = c^\dagger_{i\sigma}c^\pdagger_{i\sigma}$, and $\mu$ denotes the chemical potential. The magnitude of hopping between nearest-neighbor sites, $t$, defines the energy unit throughout this paper, $t \equiv 1 $. Note that all hopping matrix elements $t_{ij}$ will be taken to be real except for the nearest-neighbor hopping on the triangular lattice. For the latter, we use the direction-dependent values $t_{ij}=te^{\pm i\Phi/3}$  that correspond to staggered magnetic flux threading the triangular plaquettes \cite{Wang2005}, see \figref{fig:lattice}, left panel, in a way preserving lattice-translation symmetry. By changing the phase factor $\Phi$, one can systematically vary the dispersion relation, while the antiferromagnetic exchange to dominant order in $t/U$ is unaffected. Hence, the Heisenberg model and magnetic correlations associated with model (\ref{HubbardModel}) at strong coupling are independent of $\Phi$, which enters only at subleading orders in $t/U$ by inducing chiral terms \cite{PhysRevB.51.1922}. We note in passing that studying doped triangular lattice models with flux is further motivated by their relevance as minimal models of moir\'e superlattice systems \cite{ZhangSenthil,SchradeFu}, where opposite valleys have opposite flux due to time-reversal symmetry.

\begin{figure}[t]
\centering{}\includegraphics[scale=1.0]{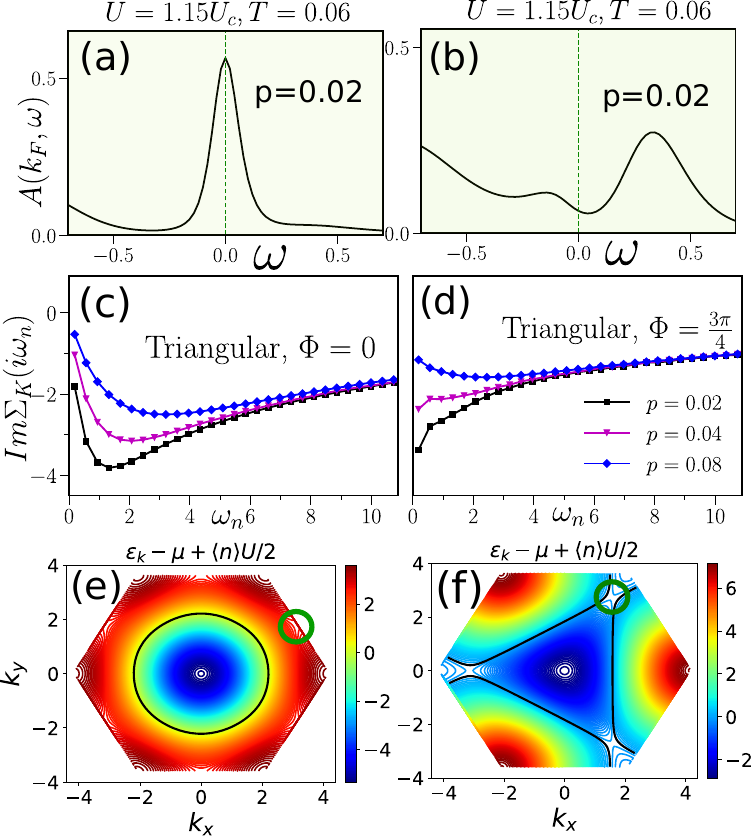}
\caption{Dependence of the PG on the bare dispersion for the triangular lattice. In (a) and (b), we show the spectral function of the triangular lattice model at low hole doping $p=0.02$ for $\Phi = 0$ [no PG, $\vec{k}_F = (1.3, 2.2)$] and  $\Phi = 3\pi/4$ [with PG, $\vec{k}_F = (0.5, \frac{2\sqrt{3}}{3}\pi)$], respectively. The imaginary part of the corresponding cluster self-energies for these values of $\vec{k}_F$ [both belong to DCA patch centered at $\vec{k}=(\frac{2\pi}{3},\frac{2\sqrt{3}}{3}\pi)$] is plotted as a function of Matsubara frequency for a few different dopings in (c) and (d). In (e) and (f), we show the Hartree-shifted dispersions $\epsilon^{*}_{\vec{k}} = \epsilon_{\vec{k}} - \mu + \braket{n}U/2$ (color plot) at $p=0.04$ together with the resulting shifted Fermi surface (black solid line), defined by $\epsilon^{*}_{\vec{k}}=0$, for $\Phi = 0$ and  $\Phi = 3\pi/4$. Furthermore, the VHSs are indicated by green circles. In all plots, we take $U=1.15U_c^{\text{Mott}}$, i.e., $U\approx 9.5$ ($U\approx 7.7$) for $\Phi = 0$ ($\Phi = 3\pi/4$).   
 \label{fig:triangle}}
\end{figure}

\sect{Triangular lattice}We first show three-site DCA results on the triangular lattice in Fig.~\ref{fig:triangle}. At half-filling, we find that the triangular lattice model without flux, $\Phi=0$, becomes insulating for sufficiently strong interactions $U > U^{\text{Mott}}_c \approx 8.2$ at $T=0.06$ --- in good agreement with previous numerical results~\cite{Hung2015,shirakawa2017,MooreZaletelTriangular}. 
In the following, we take $U=9.2$, which corresponds to $U\approx1.15\,U^{\text{Mott}}_c$ and to the regime of moderately strong correlations at which a chiral spin liquid phase was found at half-filling in \refcite{MooreZaletelTriangular} (before the spins order for $U > 10.6$). 
A characteristic feature of the PG state is the opening of a partial energy gap in the spectral function $A(\vec{k},\omega)$ near the Fermi level, at certain regions in the Brillouin zone (BZ). As can be seen in \figref{fig:triangle}(a), $A(\vec{k}_{F},\omega)$ at a typical momentum $\vec{k}_F$ on the Fermi surface [defined by $\text{Re}\,G(\vec{k}_F, \omega=0) =0$] of the slightly doped triangular lattice (hole doping $p=0.02$) with $\Phi=0$ shows a well-defined quasi-particle peak; no evidence of a PG is seen over the whole BZ. Nevertheless, when one turns on the staggered magnetic flux, the
spectral function becomes qualitatively different to its $\Phi=0$ counterpart: as can be seen in \figref{fig:triangle}(b), a PG reappears for $\Phi={3\pi}/{4}$ (same doping $p=0.02$ and even smaller $U$ to keep $U/U^{\text{Mott}}_c$ the same).

To gain further insights into this surprising difference, \figref{fig:triangle}(c) and (d) show a direct comparison of the imaginary part of the self-energy $\Sigma(\vec{k},i\omega_n)$ with $\Phi=0$ and $\Phi= {3\pi}/{4}$ for different $p$. We see that the self-energy at $\Phi =3\pi/4$ inherits an insulating-like behavior from the half-filled Mott insulator for small, finite $p$, characterized by an enhanced amplitude of scattering at small frequencies. A metallic $\Sigma(\vec{k},i \omega_n)$ is only found for significant doping ($p \geq 0.08$). In contrast, for $\Phi=0$, there is a sudden loss of low-energy scattering when the Mott insulator is doped, as the self-energy immediately turns metallic upon doping. This is in accordance with $A(\vec{k}_{F},\omega)$ in \figref{fig:triangle}(a), where no PG was observed either. Note that the absence of a PG on the triangular lattice ($\Phi=0$) was also noticed by previous numerical calculations~\cite{Kyung2007}, where the author attributes it to the triangular lattice being too frustrated. The strong sensitivity of the PG to the staggered flux shows that the criterion for the presence of a PG goes beyond the underlying nearest-neighbor Heisenberg model at half-filling, which is independent of $\Phi$.

\begin{figure}[b]
\centering{}\includegraphics[scale=0.34]{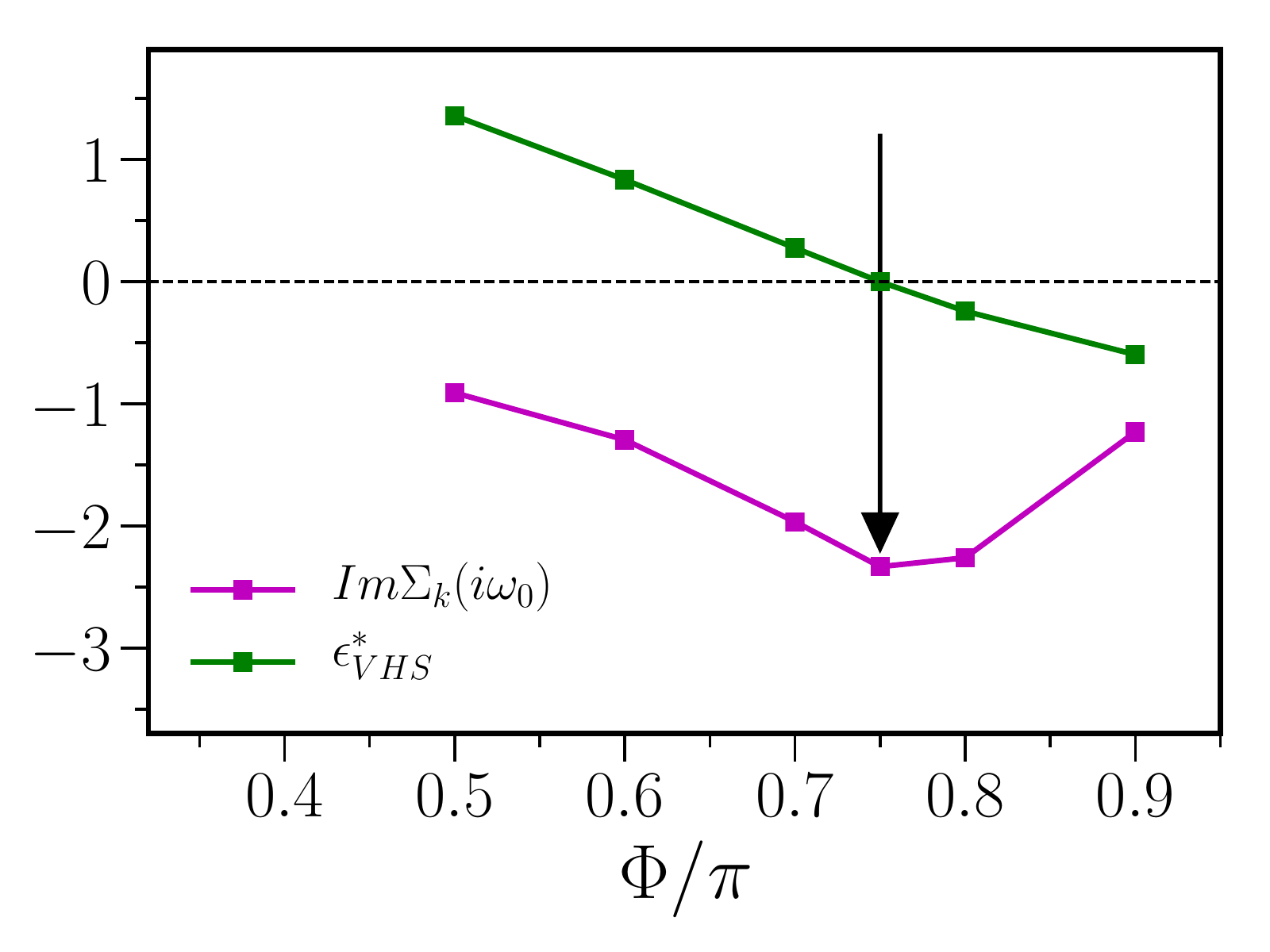}
\caption{By continuously changing the bare dispersion of the triangular lattice via $\Phi$, we find that the strongest low-energy scattering is found when the value $\epsilon^{*}_{\text{{VHS}}}$ of $\epsilon^*_{\vec{k}}$ at the momentum $\vec{k}=\vec{k}_{\text{VHS}}$ of the VHS vanishes. Here we use 3-site DCA, $U=8$, $T=0.08$, and $p=0.04$. 
\label{fig:triangle2}}
\end{figure}

\begin{figure}[b]
\includegraphics[scale=0.95]{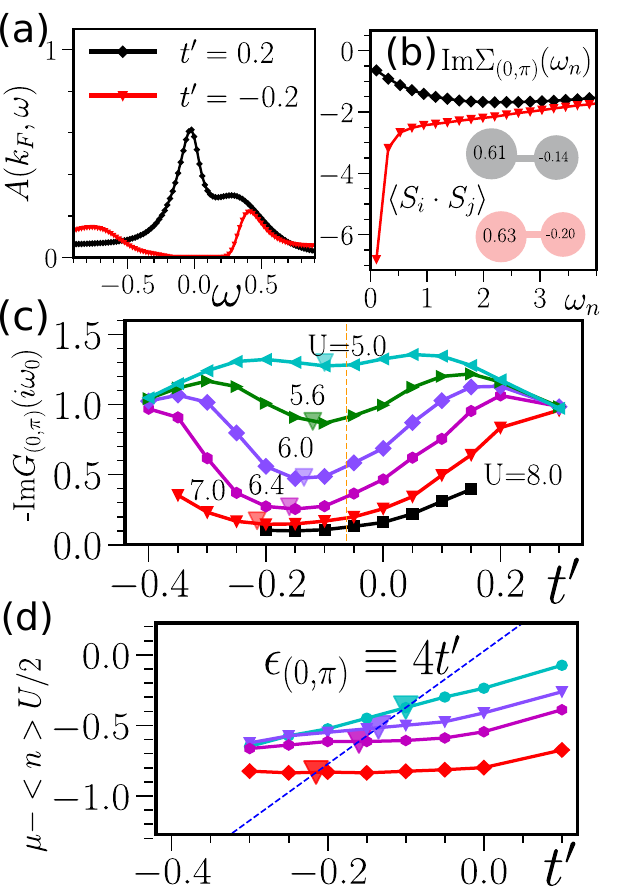}
\caption{PG and its parameter dependence on the square lattice at hole doping $p=0.05$ and $T=1/30$. (a) Spectral function at the Fermi surface for $t'=0.2$ [black line, $\vec{k}_F = (3.0,0)$] and $t'= -0.2$ [red line, $\vec{k}_F = (\pi,0)$] for $U=7$. The corresponding imaginary part of the antinodal selfenergy for the same set of parameters is shown as a function of Matsubara frequency in (b). The inset illustrates the onsite (left) and nearest-neighbor (right) spin-spin correlations using the same color code as in the main panels of (a) and (b).
In (c), the imaginary part of the antinodal Green's function at the lowest Matsubara frequency  is shown as 
a function of next-nearest neighbor hopping $t^{\prime}$ for various $U$. The minimum of the spectral intensity in (c) corresponds to the maximum of the PG. The triangles indicate the value of $t'$ at which the shifted chemical potential, solid lines in (d), is equal to $\epsilon_{(0,\pi)} = 4t'$, blue dashed line in (d), i.e., when $\epsilon^{*}_{\vec{k}} = \epsilon_{\vec{k}} - \mu + \braket{n}U/2 =0$ at the VHS $\vec{k}=(0,\pi)$.
\label{fig:square}}
\end{figure}

Since it is not the degree of geometric frustration 
which controls whether a PG is present or not, one might wonder whether there is a simple way to understand how $\Phi$ affects the PG.
To address this question, we plot, in \figref{fig:triangle}(e) and (f), the bare dispersions $\epsilon_{\vec{k}}$ with chemical potential and Hartree shift subtracted, 
\begin{equation}
    \epsilon^{*}_{\vec{k}} = \epsilon_{\vec{k}}-\mu +U\braket{n}/{2}, \label{RenormalizedDispersion}
\end{equation}
for $\Phi=0$ and  $\Phi={3\pi}/{4}$, respectively. For $\Phi=0$, the VHS, indicated by the green circle, is far away from the line where $\epsilon^{*}_{\vec{k}}=0$ (black line); note this is not the bare, non-interacting Fermi surface and not the interacting one either, which would require the self-energy in \equref{RenormalizedDispersion}. In the following, we will refer to this surface as ``shifted Fermi surface''. We observe that the VHS is located almost exactly at the shifted Fermi surface when $\Phi={3\pi}/{4}$. We also found that when $\Phi$ is continuously changed, the enhancement of low-energy scattering is essentially guided by the VHS approaching the shifted Fermi surface, as shown in Fig.~\ref{fig:triangle2}. This indicates that the position of the VHS relative to the shifted Fermi surface plays a crucial role in the formation of the PG. This observation does not only apply on the triangular lattice: on the square lattice, the enhancement of the PG is also guided by the VHS reaching the shifted Fermi surface, as we will see next.

\begin{figure*}
\begin{centering}
\includegraphics[scale=1.4]{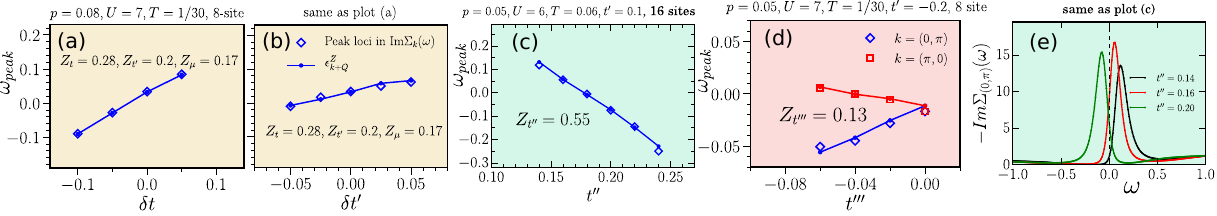}
\caption{  
In (a-d), symbols show the frequencies $\omega_{\text{peak}}$ of the peaks in Im$\Sigma_{\vec{k}}(\omega)$ [at $\vec{k}=(0,\pi)$ unless otherwise specified] found by maximum entropy method analytical continuation on $\Sigma_{\vec{k}}(i\omega_n)$, while lines display $\epsilon^{Z}_{\vec{k}+\vec{Q}} - \mu'$ with values of $Z$ marked in the corresponding plots. (a) and (b): Hubbard model with $\epsilon_{\vec{k}} = -2[(t+\delta t )\cos(k_x)+ (t-\delta t)\cos(k_y)]-4(t'+\delta t')[\cos(k_x)\cos(k_y)]$ with $t'=-0.2$, tuning $\delta t$ and $\delta t'$. (c):  $\epsilon_{\vec{k}} = -2t[\cos(k_x)+\cos(k_y)]-4t'[\cos(k_x)\cos(k_y)] - 2t''[\cos(2k_x)+\cos(2k_y)]$ with fixed $t' = 0.1$ and changing $t''$. (d):  $\epsilon_{\vec{k}} = -2t[\cos(k_x)+\cos(k_y)]-4t'[\cos(k_x)\cos(k_y)]- 2t'''[\cos(2k_x)\cos(k_y)]$ with fixed $t^{\prime} = -0.2$ and changing $t'''$  (e): Imaginary part of the antinodal selfenergy as a function of frequency. Note that while (a), (b), and (d) show 8-site DCA results, (c) and  (e) are from 16-site calculations. 
\label{fig:pole}}
\par\end{centering}
\end{figure*}

\sect{Square lattice}We now study the square lattice with nearest, $t\equiv 1$, and next-nearest neighbor hopping, $t'$, which is a prototype model for the high-$T_c$ cuprates. Following the discussion of the triangular lattice, we use the spectral function and the imaginary part of the self energy, shown in \figref{fig:square}(a) and (b), to probe the PG for two different band parameters, $t'=\pm 0.2$. 
It is worthwhile noting that the two cases $t'=\pm 0.2$ are actually identical Mott insulators at half-filling, related by a particle-hole transformation. However, upon doping, the low-energy scattering properties quickly become different: as one can see in \figref{fig:square}(a) and (b), the negative $t'$ case has a well developed PG and an insulator-like self-energy, whereas positive $t'$ lead to metallic behavior, which can, alternatively, be viewed as the asymmetry between electron- and hole-doping at fixed $t'$. This has been noticed before~\cite{civelli2005dynamical}, but the mechanism behind this striking particle-hole asymmetry has remained unclear. 

To provide a systematic approach, we show  $-\text{Im} G_{\vec{k}}(i\omega_{n=0})$ at the antinode $\vec{k}=(0, \pi)$  as a function of $t'$ for different $U$ in \figref{fig:square}(c). 
We recover that the antinodal quasiparticles are most suppressed for negative $t'$. As $U$ increases, the value of $t'$ with strongest suppression (marked by the triangles) becomes more and more negative. 
Most importantly, \figref{fig:square}(d) reveals that the same triangles also correspond to the value of $t'$ for which the VHSs [located at $\vec{k}_{\text{VHS}}=(\pi,0),(0,\pi)$] hit the shifted Fermi surface, i.e, when $\epsilon^{*}_{\vec{k}_{\text{VHS}}}=0$. In other words, just like in the triangular-lattice model, the existence of a PG on the square lattice is determined by whether a VHS is close to the line of zeros of the renormalized dispersion $\epsilon^{*}_{\vec{k}} $. In particular, this also explains the aforementioned particle-hole asymmetry regarding the PG: for $t'=-0.2$, the VHSs are closer to the shifted Fermi surface than for $t'=0.2$.

A few comments are in order. We first note that this empirical rule for the presence of a PG also applies to different values of $U$ [different colors in \figref{fig:square}(c,d)], different dopings, and holds for larger DCA cluster-sizes (16-site) or different types of hopping terms (third-nearest neighbor hopping, for example; not shown here). Let us also stress that, like on the triangular lattice, the difference between $t'=+0.2$ and $t'=-0.2$ cannot be attributed to magnetic properties: as can be seen in the inset of \figref{fig:square}(b), both the on-site and nearest-neighbor magnetic correlations are approximately the same for the two values of $t'$, although one has a metallic and the other one an insulating antinodal selfenergy. 
We finally emphasize that our criterion involves the renormalized dispersion $\epsilon^{*}_{\vec{k}}$ in \equref{RenormalizedDispersion} rather than the bare $\epsilon_{\vec{k}}$. This seems plausible as diagrammatic studies \cite{wu2017controlling} at half-filling show that the strong-coupling Feynman series converges when particle lines are expressed as $g_0^{-1}(\vec{k}, i\omega_n) = i\omega_n -\epsilon^{*}_{\vec{k}} $  and it diverges when using $g_0^{-1}(\vec{k}, i\omega_n) = i\omega_n -\epsilon_{\vec{k}}+\mu $ instead. This indicates that any perturbative treatment of PG physics should be based on $\epsilon^{*}_{\vec{k}}$ rather than on the bare $\epsilon_{\vec{k}}$. 
We emphasize, however, that the above analysis certainly cannot be perceived as a weak-coupling picture of VHS-enhanced quasiparticle scattering either: in the weak-coupling framework, the AF correlation length $\xi$ must be larger than the thermal de Broglie wave length \cite{refId0,PhysRevLett.93.147004} to obtain a PG at the ``hotspot'' of the Fermi surface. VHSs can enhance low-energy scattering but the PG continues to exist even if all VHSs are removed from the system in the weak-coupling picture. In contrast, in our strong-coupling study, $\xi$ is only a few lattice spacings and  the PG disappears in the absence of VHSs. 

\sect{Quasipole of selfenergy in PG regime}So far we have been studying the prerequisites for the existence of a PG in underdoped Mott insulators. We will next discuss the property of strong antinodal scattering when the PG is present on the square lattice. There have been various numerical studies of the square-lattice Hubbard model \cite{maier2002angle, stanescu2006fermi, wu2018pseudogap} revealing that the antinodal self-energy has a low-energy quasipole in the PG regime [see \figref{fig:pole}(e)], connected to the opening of a PG in the spectral function. 
By changing the hopping parameters, here we monitor the evolution of the quasipole of the antinodal self-energy, in the parameter regime where a strong PG is found. 
 We display results for four different variations of the hopping parameters in \figref{fig:pole} on square lattice. Remarkably, we find, that in all cases, the variation of the peak locus of the self-energy in frequency space can be described by the simple relation,
\begin{eqnarray}
 \omega_{\text{peak}}(\vec{k}) = \epsilon^{Z}_{\vec{k}+\vec{Q}} - \mu'. \label{RelationForOmegaPeak}
 \label{eq:pole}
 \end{eqnarray}
  Here $\epsilon^{Z}_{\vec{k}}$ is the bare dispersion with hopping terms renormalized by constants $Z$, and $\mu'$ is the chemical potential with Hartree shift subtracted and rescaled by $Z_{\mu}$, i.e., $\mu' = Z_{\mu}(\mu-\braket{n} U/2)$. Furthermore, $\vec{Q}= (\pi,\pi)$ is the AF wave vector. The specific values of the renormalization constants depend on the physical parameters such as doping $p$ and interaction strength $U$. 
  In \figref{fig:pole}(a) and (b), we show that, for the square lattice with nearest and next-nearest neighbor hopping,  $\omega_{\text{peak}}(\vec{k})$ for  $\vec{k}=(0,\pi)$ is described by \equref{eq:pole} with 
  $ \epsilon^{Z}_{\vec{k}}= -2 Z_t t [\cos(k_x)+\cos(k_y)] -4Z_{t'}t'[\cos(k_x)\cos(k_y)]$,  $ Z_t =0.28$, $Z_{t'}=0.2$,  and  $Z_{\mu}=0.17$.
   Note that \equref{eq:pole} is  fulfilled in all our DCA computations with different clusters and different hopping terms. 
   This is illustrated in \figref{fig:pole}(c) and (d), where a 16-site DCA on the square lattice with next-nearest ($t'$)  and third-nearest ($t''$) and an 8-site calculation with fourth-nearest neighbor hopping ($t'''$) are shown to be in accordance with \equref{eq:pole}.

While this behavior is essentially an empirical observation, the form of \equref{RelationForOmegaPeak} is motivated by a (2+1)-dimensional SU(2) gauge theory proposed for the pseudogap phase \cite{Scheurer2018,scheurer2018orbital,sachdev2019}: in that theory, the electronic self-energy is peaked at the frequencies $\omega^{\text{SU(2)}}_{\text{peak}}(\vec{k}) = \epsilon^{Z}_{\vec{k}+\vec{Q}} - \mu$, where the self-energy of the ``chargons'', the charge carriers in the bulk of the system, has poles. For instance, $Z_t \approx 0.3$ and  $Z_{t'} \approx 0.2$ were found \cite{Scheurer2018} in this description for the case of nearest and next-to-nearest neighbor hopping, in good agreement with the coefficients obtained by DCA, see \figref{fig:pole}(a) and (b).

\sect{Conclusion}In this work, we have systematically analyzed which conditions are favorable for the emergence of a PG when doping insulating half-filled Hubbard models on the triangular and square lattice. Empirically, we identified the proximity of VHSs to the shifted Fermi surface, defined by $\epsilon^*_{\vec{k}}=0$ with renormalized dispersion $\epsilon^*_{\vec{k}}$ in \equref{RenormalizedDispersion}, as favoring the formation of a PG. Theoretical clarification for why the VHSs are key in the strong-coupling regime is clearly needed. We believe that a more systematic analysis of magnetic order and potential spin-liquid phases at the half-filled triangular lattice model with staggered flux could provide important missing information to understand the observed behavior. Finally, inspired by an emergent gauge theory proposed for the pseudogap phase \cite{Scheurer2018,scheurer2018orbital,sachdev2019}, we verified the relation in \equref{RelationForOmegaPeak} for the frequency $\omega_{\text{peak}}$ at which the self-energy exhibits a quasi-pole inside the pseudogap phase for several different bare dispersions on the square lattice. 

\hspace{2em}

We thank S.~Sachdev and T.~Senthil for valuable discussions. W.~W.~acknowledges the funding from Sun Yat-sen University (Grant No. 18841204) and the National Natural Science Foundation of China (Grant No. 41030053). M.~S.~acknowledges support from the National Science Foundation under Grant No.~DMR-1664842. The Flatiron Institute is a division of the Simons Foundation.


%

\end{document}